# REAL TIME VIGILANCE DETECTION USING FRONTAL EEG

Siddarth Ganesh[1] and Ram Gurumoorthy[2]

[1]Department of Computer Science and Engineering, Amrita Vishwa Vidyapeetham, Amritapuri, Kerala, India
[2]Chief Technology Officer, StimScience Inc., Berkeley, California, USA

*ABSTRACT*

*Vigilance of an operator is compromised in performing many monotonous activities like workshop and manufacturing floor tasks, driving, night shift workers, flying, and in general any activity which requires keen attention of an individual over prolonged periods of time. Driver or operator fatigue in these situations leads to drowsiness and lowered vigilance which is one of the largest contributors to injuries and fatalities amongst road accidents or workshop floor accidents. Having a vigilance monitoring system to detect drop in vigilance in these situations becomes very important.*

*This paper presents a system which uses non-invasively recorded Frontal EEG from an easy-to-use commercially available Brain Computer Interface wearable device to determine the vigilance state of an individual. The change in the power spectrum in the Frontal Theta Band (4-8Hz) of an individual's brain wave predicts the changes in the attention level of an individual - providing an early detection and warning system. This method provides an accurate, yet cheap and practical system for vigilance monitoring across different environments.*

## 1. INTRODUCTION

Lack of vigilance leads to many accidents (on the road, in shop floors operating heavy machinery, in flying), and to drop in performance efficacy (in e-learning environments).

Fatigue and fatigue induced drowsiness have contributed to a large number of road accidents in the last few decades. While driving, the reaction time of the individuals reduces drastically with fatigue, and driving with inadequate amounts of sleep is the equivalent of driving under the influence of alcohol. The National safety council suggests that an individual is three times more likely to be in a car crash if one is fatigued while driving [1]. There have been reports stating that about 40% of the road accidents according to enforcement officers patrolling the highways and major roads are due to sleep-deprived drivers [2]. There is a critical need for non-invasive monitoring of driver vigilance to improve driving safety.

A different scenario where attention and vigilance plays a crucial role is in e-learning platforms. The attention span of individuals attending lectures online have dwindled in contrast to face-to-face lectures. This leads to a significantly lower efficacy in the learning process. Automated systems for monitoring student vigilance is essential in combating this growing problem.

Industries thrive to produce better productivity without compromising safety. With the advent of ubiquitous computing and powerful EDGE devices, shop floor monitoring for productivity analysis has increased but they have not yet been leveraged for improving the safety of operators [3]. In shop floors the exertion of an individual is already high, and compounded with the monotonous nature of the tasks the chances of drop in operator vigilance and hence machine related injuries are high.

There is a big need for simple yet accurate vigilance monitoring systems. The proposed solution can be used to identify a lack of attention or drop in vigilance in all these myriad situations.

In this paper, we present some of the related works in these different applications and the need for a simple EEG based solution. We then present the methods used for experimentation and data collection, followed by the actual data and analysis results on the efficacy of our proposed solution. We finally present a discussion of the potential applications of this method, as well as possible scope for future enhancements.

## 2. RELATED WORK

There are many solutions to track an individual's vigilance while performing different kinds of monotonous activities. Existing driver vigilance detection systems have proposed and validated the use of computer vision for relatively simple and monotonous driving tasks. The existing solutions include identifying the vigilance state using computer vision solutions - face expression, open-eye area and many more techniques complemented with deep learning to obtain inference [4]. They also extend into observing the driver's behaviour - grip on the steering wheel and movement patterns [5]. There has been some work on getting robust EEG monitoring under the low signal-to-noise environments [6]. Also in [7], under well-controlled laboratory conditions, a driver's ability to sustain attention or vigilance level is shown to be exclusively affected by these simple and monotonous driving tasks. On the other hand in real life these may not be the only contributing parameters. In addition, these systems require elaborate infrastructure and are difficult to implement in natural environments without distracting the drivers.

There have also been attempts in using software tools for gaze estimation and vibrotactile feedback systems [8] for improving attention in e-learning. There are applications developed for parents and teachers to assign, monitor and analyze the efficiency while performing assignments with the use of smartphones and eye tracking solutions [9]. There are other computer vision based posture tracking for checking student attentiveness [10]. The need for proper lighting and various other factors such as user intervention at regular intervals are really important for these systems to function as desired. There have been studies which analyse the attention of students using EEG recording devices in different contexts - for studying the change in attention to the different type of media being presented [11].

In addition, most of the current studies are based on physical behavioral changes in an individual while the physiological signals are altered much earlier than the physical behavioral expressions. Some studies suggest the use of low-cost wristbands which have the ability to record physiological signals. In [12] they have suggested the use of photoplethysmogram (PPG) and respiration signals from a sports wristband which were able to predict hypervigilance state with high accuracy indicating higher correlation of physiological sensors which do not depend on external environmental factors. Although the PPG can assist in early detection of various cardiovascular conditions, these systems are monitoring the changes in the peripheral nervous system - which are lagging indicators of the drops in vigilance compared to the changes in the central nervous system. EEG can capture these changes much earlier.

The current systems for quantitative measurement of vigilance using EEG also involve the change in reaction time and other experimental setups including simulated driving setup and simulated visual environments [13]. In [14] Oindrila et.al have implemented an amalgam of detection systems which include a single electrode (FP1) EEG recording device for driving application, combined with an infrared sensor to capture eye status (used for avoiding false positives), and a compass to detect motion in the steering wheel, to verify inactivity of the driver. These demand an elaborate infrastructure

The proposed solution is conducted in regular household setup with negligible auditory and no visual stimulation. The studies were conducted around times when vigilance is at its peak and also when it dips low for better sampling. It does not take into account reaction time of the user to the stimuli rather concentrates on identifying drops in vigilance level in real-time. The proposed solution can be utilised to monitor drivers, shop floor workers, and also students during classes non-invasively without any other environmental factor affecting the efficacy of the device or its results.

Even with the advent of EEG recording devices which can measure accurate readings at a low price, the use of cumbersome EEG recording devices with 32+ electrodes cannot be practical for most applications. The proposed solution makes use of a single frontal EEG channel (AF7) from a lightweight, non-invasive EEG headband with 5 electrodes (TP9, AF7, AF8, TP10, AUX ). The headband is made of fabric making it comfortable to be used over prolonged periods of time. It is also easy for the average individual to set up and start using the device. This headband can also be substituted with a single electrode to collect the frontal theta readings as in [14] and this data can be captured on the mobile device and alertness of the individual can be assessed.

## 3. METHODOLOGY

Four different individuals ages ranging from 14-50 participated in the protocol to analyze EEG vigilance data. The participants of the study wore a wireless EEG recording device (Muse-S headband). The EEG device connected to a mobile phone and the data was streamed from the device through BLE.

The participant was sitting on a chair in a quiet room with minimal decorations and a naturally lit environment. The participants wore the headband, had a mobile phone with a cross-platform application which was built as part of the solution. The participant was given a ten minute time period to wear the band and get accustomed to the procedure - before every session. The entire protocol took place for 3 minutes including the baseline protocol. This application was used to collect data from individuals, while reporting their vigilance status.

### 3.1. Data Acquisition

As part of the solution a cross platform application was built using Ionic Framework. The app enables the user to quickly and seamlessly connect to the MUSE-S headband using BLE (the device provides an API for this purpose). The user interface is simple and intuitive. The device and the app was used to collect a frontal channel (AF7) EEG data from the device at a 256Hz sampling rate, and data processing was done to obtain standardized values (the processing/analysis being done is explained in the next section).

The application collects data to obtain baseline values when the participant is calm and relaxed. A vigilance threshold is used to evaluate the vigilance state of the individual (as explained in the next section).

Data collection sessions for each individual were done at 3 different times throughout the day - in the morning when the mind and body is active, post lunch when the alerting signal of the circadian rhythm gradually dips around 1.30 PM, and finally when the build-up of adenosine is at its peak right before bedtime around 9.30 PM [15]. This was done to make sure that the monitor is accurate across the different times and circadian rhythmic state of the individual.

We also recorded 2 sessions at each of these 3 time periods : the first session included instructions on when the individual had to close their eyes and when they had to open at regular intervals, this status was auto tagged using the mobile application. For the second session the subjects were allowed to naturally keep their eyes open or closed, and the experimenter tagged their eye status using the app.

### 3.2. Data Processing (Analysis)

The data from the collected EEG was processed by calculating the PSD (power spectral density) using DFT (Discrete Fourier Transform). Then we cumulated the spectrum in the theta band range (4-8Hz). This procedure is repeated every 5 seconds and the Theta band power (Theta BP) is obtained.

We had the subjects have their eyes closed for the first 30 seconds, and used that duration to obtain a baseline for their individual Theta BP when they are non-vigilant. We then used that mean Theta BP (across the six 5 second epochs) in setting a threshold for estimating vigilance.
The remaining data collection epochs were then classified as the subject being vigilant or non-vigilant using this threshold (by using a standard scaling of the mean BP - here 1.1). This scaling factor was selected after looking at a few subject sessions to optimize prediction accuracy, and was then used across the data sessions.

The data from the remaining 2 minutes and 30 seconds of the session was used to test the prediction of vigilance state using the threshold limit. The BP for every 5 seconds is collected, calculated and compared with the threshold BP of that individual and is used to determine if the individual is in a non-vigilant state. If Theta BP is over the threshold (the frontal theta power increased) they are vigilant, and vice versa.

## 4. RESULTS

The EEG data is received and processed by a mobile phone and the user's vigilance state is indicated in real-time.

Figure 1 presents a chart of the Theta Band Power across epochs (across 5s windows) for a participant, that is cumulated when they had their eyes closed and when they had their eyes open. As the chart shows, there is a good separation between these 2 data sets - which indicates that the Theta Band Power could be a good criterion for predicting the vigilance state of an individual.

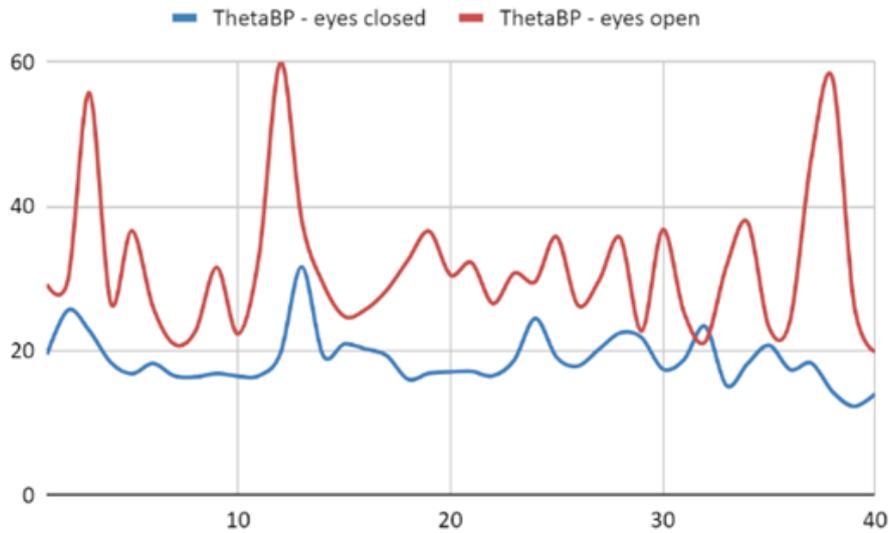

Figure 1. Theta Band Power Variations

When we look at using this criterion for predicting the eyes closed/open state (vigilant/non-vigilant state), the prediction accuracy across multiple sessions and multiple individuals has been tabulated in Table 1.

Table 1. Session Wise Accuracy.

| Session ID | Instructed | Natural | Combined |
|---|---|---|---|
| P1-Morning | 84.85% | 87.88% | 86.36% |
| P1-Afternoon | 90.91% | 93.94% | 92.42% |
| P1-Night | 84.85% | 90.91% | 87.88% |
| P2-Morning | 84.85% | 93.94% | 89.39% |
| P2-Afternoon | 90.91% | 87.88% | 89.39% |
| P2-Night | 90.91% | 93.94% | 92.42% |
| P3-Morning | 93.94% | 87.88% | 90.91% |
| P3-Afternoon | 84.85% | 96.97% | 90.91% |
| P3-Night | 85.71% | 96.97% | 91.34% |
| P4-Morning | 87.88% | 93.94% | 90.91% |
| P4-Afternoon | 84.85% | 84.85% | 84.85% |
| P4-Night | 87.88% | 81.82% | 84.85% |
| Average | 87.70% | 90.91% | 89.30% |
| STD | 3.23% | 4.83% | 4.34% |

The sessions where the participants were explicitly instructed to close and open their eyes at specified intervals had relatively lesser accuracy compared to the sessions in which they tagged themselves while they closed or opened their eyes. Across the sessions the natural baselining protocol had a prediction accuracy of 91% and the instructed mode had a slightly lower prediction accuracy of 88%. But when we do a T-test of the 2 sets of data (instructed vs natural estimation accuracy) we see the p-value is 0.098. Though a value of 0.05 is usually used for rejecting the null hypothesis, 0.10 is still seen as indistinguishable (still both sets are considered to be from the same underlying population).

In the data, we can see that in Table 2, the off-diagonal accuracy (closed eye condition being predicted as open) is a little high (over 20%). We estimate that it might be an artifact of instructing the participants to close/open their eyes. When they have been closed and we instruct them to open their eyes, there may be a few seconds before they really get vigilant. This may be contributing to this higher error. We also see that in the natural condition (Table 3) where they automatically close/open eyes, this off diagonal accuracy (closed eye condition being predicted as open) drops to around 11% (half of the error rate from the instructed condition) - supporting the above hypothesis.

Table 2. Confusion matrix of actual vs estimated eye status – Instructed Method

| Actual\Estimated | Closed | Open |
|---|---|---|
| Closed | 76.79% | 4.39% |
| Open | 23.21% | 95.61% |

Table 3. Confusion matrix of actual vs estimated eye status – Natural Method

| Actual\Estimated | Closed | Open |
|---|---|---|
| Closed | 88.38% | 8.12% |
| Open | 11.62% | 91.88% |

## 5. CONCLUSIONS AND DISCUSSIONS

The system shows that the vigilance (alertness) state of an individual can be well predicted by using a single channel wireless EEG data. The simplicity of just using a frontal EEG (on the forehead) makes the system quite non-invasive and usable in many environments without encumbering or distracting the operators.

Figure 2 presents a flow chart of using this system for predicting vigilance in any application. One example could be driving, where the dashboard could be linked to this system to display a danger indicator and/or play an alert sound when the user is in a non-vigilant state.

Another example could be in the use of this system to alert floor managers in manufacturing applications to alert the non-vigilant state of heavy-machinery-operators. This could bring down the hazards and accidents. It can also be used for student attention monitoring in remote learning situations as discussed in the introduction, or in pilot vigilance monitoring.

One point of discussion is the window length (time duration) used for the PSD and band power estimation. Typically people use anywhere between 2s-10s windows for this. The lower the window length, the larger the noise in the estimate but more responsiveness in the alertness monitor. The larger the window length, the lower the noise in the estimate but more sluggish in the alertness monitoring. We have chosen a balance between the ranges and selected 5 seconds for our estimation. This window can be further optimized by the application for which it is used - as it can dictate the responsiveness required and the acceptable noise rate.

In the future we want to explore a couple of areas: we want to expand the data collection, with more participants and over longer time durations and environments; we also would like to use machine learning models for predicting the vigilance state (using the raw electrode signal and/or power spectral density of the signal) - we anticipate that these models could further improve the accuracy of the prediction.

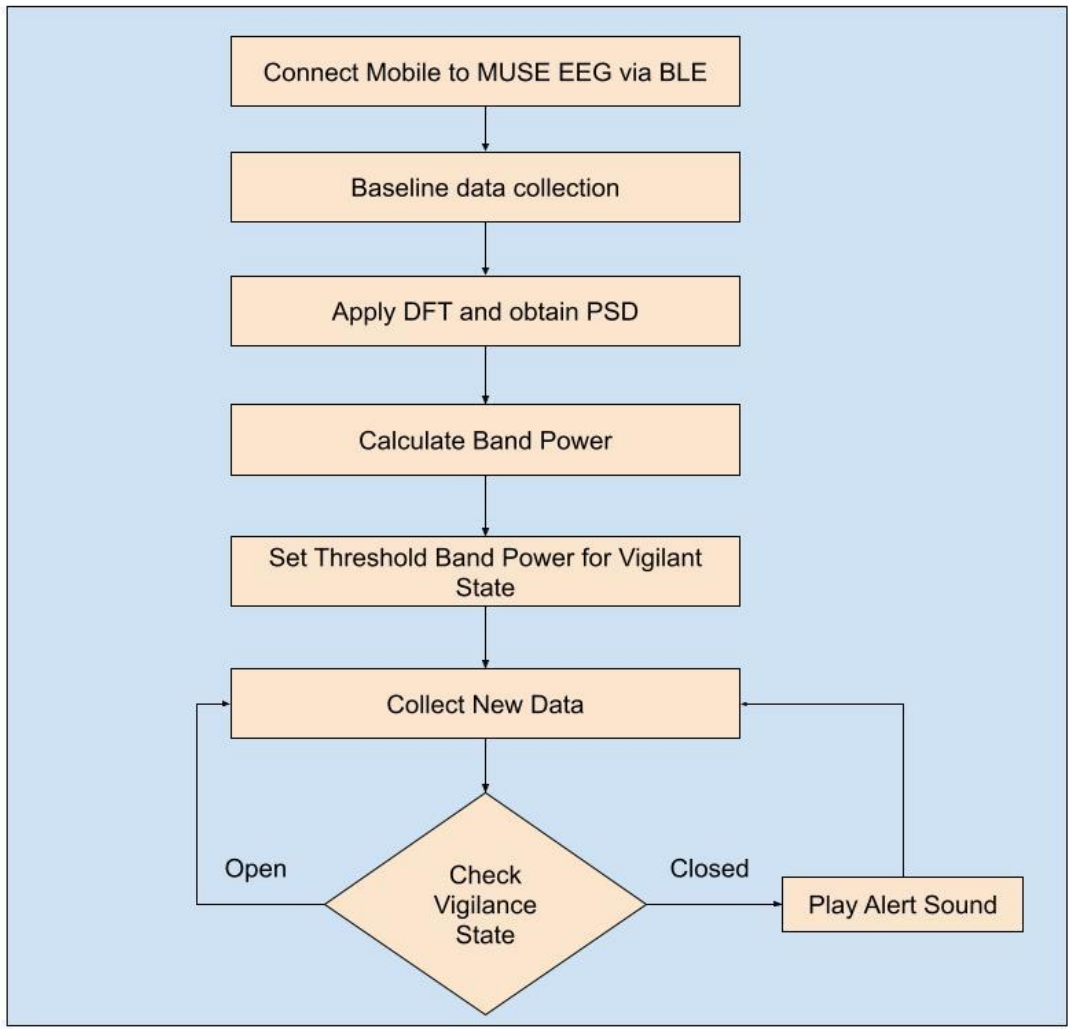

Figure 2. System flowchart for predicting vigilance

## REFERENCES

[1] "National safety council" [Online]. Available: https://www.nsc.org/road/safety-topics/fatigued-driver

[2] "Sleep-deprived drivers responsible for 40% of road accidents, say transport officials." [Online]. Available: https://www.thehindu.com/news/national/kerala/sleep-deprived-drivers-responsible-for-40-of-road-accidents-say-transport-officials/article30868895.ece

[3] Mourtzis, Dimitris & Milas, Nikolaos & Vlachou, Katerina. (2018). An Internet of Things-Based Monitoring System for Shop-Floor Control. Journal of Computing and Information Science in Engineering. 18. 021005. 10.1115/1.4039429.

[4] M. Garćıa-Garćıa, A. Caplier, and M. Rombaut, "Sleep deprivation detection for real-time driver monitoring using deep learning," in Image Analysis and Recognition, A. Campilho, F. Karray, and B. terHaar Romeny, Eds.Cham: Springer International Publishing, 2018,pp. 435–442.

[5] Eskandarian, Azim & Mortazavi, Ali. (2007). Evaluation of a Smart Algorithm for Commercial Vehicle Driver Drowsiness Detection. IEEE Intelligent Vehicles Symposium, Proceedings. 553 - 559. 10.1109/IVS.2007.4290173.

[6] H. Yu, H. Lu, T. Ouyang, H. Liu, and B. Lu, "Vigilance detection based on sparse representation of eeg," in 2010 Annual International Conference of the IEEE Engineering in Medicine and Biology, 2010,pp. 2439–2442.

[7] Z. Guo, Y. Pan, G. Zhao, S. Cao, and J. Zhang, "Detection of driver vigilance level using eeg signals and driving contexts,"IEEE Transactions On Reliability, vol. 67, no. 1, pp. 370–380, 2018.

[8] B. Thankachan, "Haptic feedback to gaze events," Ph.D. dissertation, 122018.

[9] M. A. Nuño-Maganda, C. Torres-Huitzil, Y. Hernández-Mier, J. De LaCalleja, C. C. Martinez-Gil, J. H. B. Zambrano, and A. D. Manríquez,"Smartphone-based remote monitoring tool for e-learning,"IEEE Ac-cess, vol. 8, pp. 121 409–121 423, 2020.

[10] A. Revadekar, S. Oak, A. Gadekar, and P. Bide, "Gauging attention of students in an e-learning environment," in 2020 IEEE 4th Conference on Information Communication Technology (CICT), 2020, pp. 1–6.

[11] D. Ni, S. Wang, and G. Liu, "The eeg-based attention analysis in multimedia m-learning," Computational and Mathematical Methods inMedicine, vol. 2020, pp. 1–10, 06 2020.

[12] B. Lee, B. Lee and W. Chung, "Wristband-Type Driver Vigilance Monitoring System Using Smartwatch," in IEEE Sensors Journal, vol. 15, no. 10, pp. 5624-5633, Oct. 2015, doi: 10.1109/JSEN.2015.2447012.

[13] C. Lin, C. Chuang, C. Huang, S. Tsai, S. Lu, Y. Chen, and L. Ko,"Wireless and wearable eeg system for evaluating driver vigilance,"IEEE Transactions on Biomedical Circuits and Systems, vol. 8, no. 2,pp. 165–176, 2014.

[14] O. Sinha, S. Singh, A. Mitra, S. K. Ghosh, and S. Raha, "Development Of a drowsy driver detection system based on eeg and ir-based eye blink detection analysis," in Advances in Communication, Devices and Networking, R. Bera, S. K. Sarkar, and S. Chakraborty, Eds. Singapore:Springer Singapore, 2018, pp. 313–319.

[15] "External factors that influence sleep." [Online]. Available: http://healthysleep.med.harvard.edu/healthy/science/how/external-factors